\documentclass{article}
\usepackage{arxiv}

\usepackage[utf8]{inputenc} 
\usepackage[T1]{fontenc}    
\usepackage{hyperref}       
\usepackage{url}            
\usepackage{booktabs}       
\usepackage{amsfonts}       
\usepackage{nicefrac}       
\usepackage{microtype}      
\usepackage{lipsum}
\usepackage{amsmath,mathtools,bbm,multirow,url,hyperref,enumerate}
\usepackage{color}
\usepackage{float}
\usepackage{hyperref}
\usepackage{makecell}
\usepackage{algorithm,algorithmic}
\usepackage{multicol}

\title{Combining cytotoxic agents with continuous dose levels in seamless phase I-II clinical trials}

\author{
  Jos\'e L. Jim\'enez \\
  Novartis Pharma A.G.\\
  Basel, Switzerland \\
  \texttt{jose\_luis.jimenez@novartis.com} \\
    \And
  Mourad Tighiouart \\
  Cedars-Sinai Medical Center\\
  Los Angeles, USA \\
  \texttt{Mourad.Tighiouart@cshs.org}
}

\begin{document}
\maketitle

\begin{abstract}
Phase I-II cancer clinical trial designs are intended to accelerate drug development. In cases where efficacy cannot be ascertained in a short period of time, it is common to divide the study in two stages: i) a first stage in which dose is escalated based only on toxicity data and we look for the maximum tolerated dose (MTD) set and ii) a second stage in which we search for the most efficacious dose within the MTD set. Current available approaches in the area of continuous dose levels involve fixing the MTD after stage I and discarding all collected stage I efficacy data. However, this methodology is clearly inefficient when there is a unique patient population present across stages. In this article, we propose a two-stage design for the combination of two cytotoxic agents assuming a single patient population across the entire study. In stage I, conditional escalation with overdose control (EWOC) is used to allocate successive cohorts of patients. In stage II, we employ an adaptive randomization approach to allocate patients to drug combinations along the estimated MTD curve, which is constantly updated. The proposed methodology is assessed with extensive simulations in the context of a real case study.
\end{abstract}

\section{Introduction}
\label{sc_introduction}

Integrated phase I-II clinical trial designs are efficient approaches to accelerate drug development. This seamless process uses a single protocol to investigate two aspects of the development of a compound, namely safety and efficacy.

With the use of chemotherapy compounds, a dose limiting toxicity (DLT) is generally ascertained after one cycle of therapy for the purpose of estimating the maximum tolerated dose (MTD). If the efficacy outcome can be ascertained in a short period of time (e.g., after one or two cycles of therapy), it is common to employ one-stage designs where the joint probability of toxicity and efficacy is sequentially updated after each cohort of patients. Extensive methodology to carry out such designs can be found in  \cite{thall1998strategy, braun2002bivariate, ivanova2003new, thall2004dose, chen2015dose, sato2016adaptive, yuan2009bayesian} for single agent trials and in \cite{huang2007parallel, yin2006bayesian} for drug combinations. In contrast, when efficacy cannot be ascertained in a short period of time, we may employ two-stage designs where a set of maximum tolerated dose combinations are first selected, and then tested for efficacy in a subsequent second stage (see e.g., \cite{rogatko2008patient, le2009dose, chen2012methodology, shimamura2018two, yuan2011bayesian, zhang2016practical,tighiouart2019two,jimenez2020bayesian}).

This work is motivated by the cisplatin-cabazitaxel clinical trial design proposed by \cite{tighiouart2019two}. In that phase I-II study design, stage I was inspired by a previous phase I trial published by \cite{lockhart2014phase} in patients with advanced solid tumors, where a single MTD of cisplatin/cabazitaxel 15/75 $mg/m^2$ was recommended. However, based on their results and other preliminary efficacy data, it was hypothesized by the clinical team that there could be a series of dose combinations which could be tolerable and efficacious in prostate cancer. In stage II of \cite{tighiouart2019two}, the design enrolls 30 additional patients to identify dose combinations with high probability of efficacy, along the MTD curve estimated from the first stage data.

A limitation of the original cisplatin-cabazitaxel design by \cite{tighiouart2019two} is that uncertainty about the estimated MTD curve is not taken into account in stage II. Moreover, efficacy data from stage I were not used to model the dose-efficacy relationship in stage II. The rational for this disconnection was that stage I and stage II populations were not exactly the same, and therefore it was hypothesised that the dose-efficacy profiles could also be different. Recently, \cite{jimenez2021bayesian} proposed a method to robustly combine efficacy data across stages in this particular setting. In practice however, it is not uncommon to design phase I-II clinical trials with the same patient population across stages. In such a situation, it would be more sensible to use all the toxicity and efficacy data collected from stage I as well as update the MTD curve during stage II. To the best of our knowledge, such a design does not exist in the phase I-II clinical trial designs literature with combination of cytotoxic agents and continuous dose levels.

In this article, we extend the work of \cite{tighiouart2019two} by allowing the use of all the safety and efficacy data collected throughout the entire trial, which implies that the MTD curve is sequentially updated during stage II. Since we assume that efficacy cannot be ascertained in a short period, the design proceeds in two stages. The design's operating characteristics are evaluated by extensive simulations in the context of the cisplatin-cabazitaxel clinical trial.

The rest of the manuscript is organized as follows. In section \ref{sc_motivating_example} we describe the clinical trial that motivates this research. In section \ref{sc_method} we introduce the dose-toxicity and dose-efficacy models as well as the study design. In section \ref{sc_simulation_study} we present the operating characteristics under several scenarios for the true dose-toxicity and dose-efficacy relationships. Last, in section \ref{sc_discussion}, we discuss the major contributions of this manuscript together with potential areas where there may be room for improvement.

\section{The cisplatin-cabazitaxel trial}
\label{sc_motivating_example}

The two-stage phase I-II cisplatin-cabazitaxel trial design proposed by \cite{tighiouart2019two} investigates the combination of cisplatin and cabazitaxel, given every 3 weeks, in patients with metastatic, castration resistant prostate cancer. A continuous set of doses that range from 10 to 25 $mg/m^2$ for cabazitaxel and from 50 to 100 $mg/m^2$ for cisplatin were administered intravenously. In stage I, the study enrolls 30 patients using conditional escalation with overdose control (EWOC) algorithm (see \cite{tighiouart2017bayesian}) to estimate the MTD curve. In stage II, the study enrolls another 30 patients from the same population of patients but with visceral metastasis to identify dose combinations with high probability of efficacy along the MTD curve estimated at the end of stage I. These patients are allocated to dose combinations along the MTD curve using a Bayesian adaptive design after modeling the dose-efficacy curve with cubic splines.

The first stage of the cisplatin-cabazitaxel trial was motivated by the phase I trial published by \cite{lockhart2014phase}, where the combinations of cisplatin/cabazitaxel 15/75 $mg/m^2$, 20/75 $mg/m^2$ and 25/75 $mg/m^2$ were explored in patients with advanced solid tumors, using a ``3+3'' design. The recommended MTD was 15/75 $mg/m^2$ on the basis of data from 9 evaluable patients in phase I and 15 patients in the expansion cohort, although only 2 patients out of 18 treated at the recommended MTD had DLT. Based on this low toxicity rate and preliminary efficacy data, the investigators hypothesized that a series of tolerable dose combinations which could be efficacious exist in prostate cancer.

The original cisplatin-cabazitaxel trial design was tailored for a situation with different patient populations across stages. As previously mentioned, it is not uncommon to have the same patient population across the entire study. Under this assumption, having two disconnected stages would not be sensible since the MTD could be constantly updated and the efficacy data collected in stage I could be used in stage II. Therefore, we propose an extension of the original cisplatin-cabazitaxel trial design to a situation where we have either the same patient population across the two stages or very similar populations with respect to their susceptibility to toxicities and clinical benefit, the primary response endpoint in the cisplatin-cabazitaxel trial.

\section{Method}
\label{sc_method}
\subsection{Formulation of the problem}
\label{sc_problem_formulation}

Let $\{X_{\min}, Y_{\min}, X_{\max}, Y_{\max}\}$ be the lower and upper bound of the continuous dose levels $x$, $y$ of the compounds $X$ and $Y$, respectively. The doses are standardized to fall within the interval [0,1] using the transformations $h_1(x) = (x - X_{\min}) / (X_{\max} - X_{\min})$ and $h_2(y) = (y - Y_{\min}) / (Y_{\max} - Y_{\min})$, so that $x \in [0,1]$ and $y \in [0,1]$.

Let $Z = \{0,1\}$ be the binary indicator of DLT where $Z = 1$ represents the presence of a DLT after a predefined number of treatment cycles, and $Z = 0$ otherwise. Let $E = \{0,1\}$ be the indicator of treatment response where $E = 1$ represents a positive response after a predefined number of treatment cycles, and $E = 0$ otherwise. Following Tighiouart (2019) \cite{tighiouart2019two}, let $\theta_Z = 0.33$ be the target probability of DLT and $\theta_E = 0.15$ be the probability of efficacy of the standard of care treatment. The goal of this phase I-II trial is to identify dose combinations $(x,y)$ with probability of DLT $\pi_Z(x,y) \leq \theta_Z$ and probability of efficacy $\pi_E(x,y) \geq \theta_E$. When employing synergistic cytotoxic agents, it is common to assume that both the dose-toxicity and dose-efficacy relationship are monotonically increasing functions. This implies that the optimal dose will lie in the MTD set, defined as $\mathcal{M} = \left \{ (x,y): \pi_Z(x,y) = \theta_Z \right \}$.

\subsection{Definition of the marginal dose-toxicity and dose-efficacy models}
\label{sc_marginal_models}

Following the work of \cite{ivanova2009adaptive,cai2014bayesian,lyu2019aaa}, we assume that the binary outcomes of toxicity and efficacy are independent. Let the marginal probability of DLT and marginal probability of efficacy models be defined as

\begin{equation}
\label{eq_prob_dlt}
    \pi_Z(x,y) = P(Z = 1 | x, y) = F(\alpha_0 + \alpha_1 x + \alpha_2 y + \alpha_3 x y), 
\end{equation}and
\begin{equation}
\label{eq_prob_eff}
 \pi_E(x,y) = P(E = 1 | x, y) = F(\beta_0  + \beta_1 x + \beta_2 y + \beta_3 xy + \beta_4 x^2 + \beta_5 y^2),
\end{equation}respectively, with $\alpha_1,\alpha_2,\alpha_3,\beta_1,\beta_2,\beta_3 > 0$, and where $F(.)$ represents the cumulative distribution function of the standard normal distribution. Note that in (\ref{eq_prob_dlt}), DLT is assumed to be attributed to either drug $X$ or $Y$, or both, i.e.,  we do not take into account toxicity attributions (see \cite{jimenez2019cancer}) because the number of attributable DLTs is expected to be very low given the cytotoxic nature of cisplatin and cabazitaxel.

In (\ref{eq_prob_eff}), we added quadratic terms $\beta_4 x^2, \beta_5 y^2$ to allow more flexibility in the marginal efficacy model $\pi_E(x,y)$. The rationale for this is that the dose-efficacy profiles along the MTD curve can become quite complex (see Figure \ref{Figure_2}) when having synergistic compounds (i.e., $\alpha_3 > 0$) even if we only consider linear terms in both marginal models. If we also account for the uncertainty in the MTD curve estimation and the relatively small sample size available, in the development of this research we have found that a marginal efficacy model $\pi_E(x,y)$ with only linear terms was, in certain situations, not able to capture the dose-efficacy profile along the MTD curve. In our simulation scenarios, the optimal dose combination is guaranteed to be on the MTD curve because the underlying true dose-efficacy relationship (i.e., the data-generating model) is a monotonically increasing function. However, in a setting with dose-toxicity profiles without drug-drug interactions (i.e., $\alpha_3 = 0$), this additional flexibility would not be necessary and a marginal efficacy model with only linear terms would be sufficient to capture any dose-efficacy profile along the MTD curve.

Following \cite{tighiouart2017bayesian}, we reparameterize the marginal probability of DLT defined in equation \eqref{eq_prob_dlt} in terms of parameters that clinicians can easily interpret. Let $\rho_{uv}$ denote the probability of DLT when the levels of agents $X = u$ and $Y = v$, with $u = 0, 1$, and $v = 0, 1$, so that $\alpha_0 = F^{-1}(\rho_{00})$, $\alpha_1 = (F^{-1}(\rho_{10}) - F^{-1}(\rho_{00}))$, and $\alpha_2 = (F^{-1}(\rho_{01}) - F^{-1}(\rho_{00}))$. It then follows that the MTD set takes the form

\begin{equation}
\label{mtdcurve}
\begin{split}
& \mathcal{M} = \left \{ (x,y): y = \frac{(F^{-1}(\theta) - F^{-1}(\rho_{00})) - (F^{-1}(\rho_{10}) - F^{-1}(\rho_{00})) x}{(F^{-1}(\rho_{01}) - F^{-1}(\rho_{00})) + \alpha_3 x } \right \}. 
\end{split}
\end{equation}

\subsection{Likelihood and posterior distributions}

We assume that $\rho_{10}, \rho_{01}$ and $\alpha_3$ are independent \emph{a priori} with $\rho_{01} \sim \mbox{beta}(1,1)$, $\rho_{10} \sim \mbox{beta}(1,1)$, and conditional on $(\rho_{01}, \rho_{10})$, $\rho_{00} / \min (\rho_{01}, \rho_{10}) \sim \mbox{beta}(1,1)$. Also, let the interaction parameter $\alpha_3 \sim \mbox{gamma(0.1, 0.1)}$. Let $D_{n} = \{(x_i,y_i,Z_i, E_i), i= 1, \dots, n\}$ be the data collected after $n$ patients have been enrolled. The posterior distribution of the dose-toxicity model parameters is

\begin{equation*}
\begin{split}
\pi(\rho_{00}, \rho_{10}, \rho_{01}, \alpha_3 | D_n) & \propto
\prod_{i=1}^{n} G((\rho_{00}, \rho_{10}, \rho_{01}, \alpha_3; x_i, y_i))^{Z_i} \times (1 - G(\rho_{00}, \rho_{10}, \rho_{01}, \alpha_3; x_i, y_i))^{1-Z_i} \\
& \times \pi(\rho_{01}) \pi(\rho_{10}) \pi(\rho_{00} | \rho_{01},\rho_{10}) \pi(\alpha_3),
\end{split}
\end{equation*}where
\begin{equation*}
\begin{split}
G(\rho_{00}, \rho_{10}, \rho_{01}, \alpha_3; x_i, y_i) & = F(F^{-1}(\rho_{00}) + (F^{-1}(\rho_{10}) - F^{-1}(\rho_{00})) x_i  \\
& + (F^{-1}(\rho_{01}) - F^{-1}(\rho_{00})) y_i + \alpha_3 x_i y_i).
\end{split}
\end{equation*}

In the dose-efficacy model, we do not incorporate any prior knowledge. Thus, we assume that $\beta_0, \beta_4, \beta_5 \sim \mbox{N}(0, 10^2)$, $\beta_1, \beta_2, \beta_3 \sim \mbox{Gamma}(0.1, 0.1)$ and that $\beta_j, j=0,\ldots,5$ are independent \emph{a priori}. The posterior distribution of the dose-efficacy model parameters is 

\begin{equation*}
\begin{split}
\pi(\beta_0, \beta_1, \beta_2, \beta_3, \beta_4, \beta_5 | D_n) & \propto \prod_{i=1}^{n} \pi_E(x_i, y_i)^{E_i} \times (1 - \pi_E(x_i, y_i))^{1-E_i} \\
& \times \pi(\beta_0) \pi(\beta_1) \pi(\beta_2) \pi(\beta_3) \pi(\beta_4) \pi(\beta_5)
\end{split}
\end{equation*}

All Bayesian computations are done using \texttt{R} (see \cite{r2019}) and \texttt{JAGS} (see \cite{plummer2003jags}).

\subsection{Study design}
\label{sc_study_design}

Stage I will enroll a total of $N_1 = C_1 \times m_1$ patients, where $C_1$ represents the total number of cohorts in phase I with equal number of patients $m_1$. Stage II will enroll a total of $N_2 = C_2 \times m_2$ patients. Let $N = N_1 + N_2$ be the total number of patients that the entire study will enroll and $\widehat{\mathcal{M}}_{N}$ be the estimated MTD set based on data from all $N$ patients. At the end of stage II, we test the following null and alternative hypotheses
\begin{equation}
\label{eq_hypotheses}
\begin{split}
& H_0: \pi_E(x,y) \leq \theta_E \text{ for all  } (x,y) \in \widehat{\mathcal{M}}_{N} \quad \text{vs.} \\
& H_1: \pi_E(x,y) > \theta_E \text{ for some  } (x,y) \in \widehat{\mathcal{M}}_{N},
\end{split}
\end{equation}and we reject the null hypothesis if 
\begin{equation}
    \underset{(x,y) \in \widehat{\mathcal{M}}_{N}}{\mbox{max}} P(\pi_E(x,y) > \theta_E | D_N) > \delta_u,
    \label{eq_optimal_dose_comb}
\end{equation} where $\delta_u$ is a pre-specified design parameter. Moreover, the dose combination
\begin{equation}
\label{opt_dose}
(x,y)_{opt}=\underset{(x,y) \in \widehat{\mathcal{M}}_{N}}{\mbox{argmax}} P(\pi_E(x,y) > \theta_E | D_N), 
\end{equation}is recommended as the optimal dose combination and is selected for further phase II or III studies.

Stage I is based on the escalation with overdose control (EWOC) principle (see \cite{babb1998cancer, tighiouart2005flexible, tighiouart2010dose, tighiouart2017bayesian, tighiouart2012number, shi2013escalation}) where the posterior probability of overdosing the next cohort of patients is bounded by a feasibility bound $\alpha$. In a cohort with two patients, the first one would receive a new dose of compound $X$ given that the dose $y$ of compound $Y$ that was previously assigned. The other patient would receive a new dose of compound $Y$ given that dose $x$ of compound $X$ was previously assigned. These steps are described in Stage I of Algorithm \ref{algorithm1}. Using EWOC, these new doses are at the $\alpha$-th percentile of the conditional posterior distribution of the maximum tolerated dose combinations. The feasibility bound $\alpha$ increases from 0.25 up to 0.5 in increments of 0.05 (see \cite{wheeler2017toxicity}).Accrual continues until the maximum sample size in stage I is reached or the trial is stopped early for safety.

Stage II follows the response-adaptive randomization principle. This type of Monte Carlo algorithm uses the current parameter estimates to sample a cohort of $m_2$ dose combinations from the estimated dose-efficacy standardized density of $\widehat{\pi}_E (x,y)$ along the estimated MTD curve. Note that $\widehat{\pi}_E (x,y)$ uses the Bayes estimates of the dose-efficacy model parameters. Because stage II selects doses on the estimated MTD curve $\widehat{\mathcal{M}}$, and there is a one-to-one correspondence between $(x,y) \in \widehat{\mathcal{M}}$ and $x \in X'$, where $X' \subset X$, we may write $\widehat{\pi}_E (x,y) = \widehat{\pi}_E (x)$ for $(x,y) \in \widehat{\mathcal{M}}$. The standardized density of  the estimated efficacy curve is $\widehat{\pi}_E^{\tiny \mbox{STDZ}} (x) = \frac{\widehat{\pi}_E (x)}{\int_{x \in X' \widehat{\pi}_E (x) dx}}$. A rejection sampling algorithm is then used to sample $m_{2}$ dose combinations from this density.

These steps are are described in Stage II of Algorithm \ref{algorithm1}. Note that in this stage, the MTD curve is updated after the DLT status of each cohort of $m_2$ patients is resolved and a new cohort of $m_2$ patients are randomly selected from the new MTD curve according to the standardized density $\widehat{\pi}_E (x,y)$.

Another contribution of this article with respect to \cite{tighiouart2019two,jimenez2020bayesian,jimenez2021bayesian} is that we do not require the first cohort of stage II to be homogeneously distributed along the estimated MTD curve. The justification is that, in this article, we do not have two disconnected stages and thus we use all the available stage I efficacy data to already have posterior estimates of $\beta_0, \dots, \beta_5$ at the end of stage I as displayed in line 13 of Algorithm \ref{algorithm1}. This allows to allocate the first cohort of patients in stage II to dose combination following the adaptive randomization principle.

To facilitate the understanding of Algorithm \ref{algorithm1}, we  define the variables that will be used. In stage I, $c_1$ denotes the current cohort indicator with each cohort containing 2 patients. For example, $c_1 = 2$ (i.e., the second cohort) contains the dose combinations from patients 3 and 4. In a generalized way, one cohort contains patients $(2 c_1 - 1)$ and $(2 c_1)$ and so, if $c_1 = 2$, we know that the second cohort contains the patients $2 c_1 - 1 = 3$ and $2 c_1 = 4$. Now let $(x_{2 c_1 - 1}, y_{2 c_1 - 1})$ and $(x_{2 c_1}, y_{2 c_1})$ denote the dose combinations to which the two patients from cohort $c_1$ were (or will be) allocated. Using the same example, if $c_1 = 2$, we see that $(x_{2 c_1 - 1}, y_{2 c_1 - 1}) = (x_3, y_3)$ and $(x_{2 c_1}, y_{2 c_1}) = (x_4, y_4)$ correspond to the dose combinations from patients 3 and 4. Also, let $\pi(\Gamma_{X|Y=y} | D_n)$ represent the posterior distribution of the MTD of drug $X$ given that the level of drug $Y$ is equal to $y$ (i.e., given that $Y$ is fixed) and given the data $D_n$ (see equation \eqref{mtdcurve} for the definition of the MTD). Last, let $\Pi_{\Gamma_{X | Y = y}}^{-1}(\alpha | D_n)$ denote the $\alpha$-th percentile of $\pi(\Gamma_{X|Y=y} | D_n)$.

\begin{algorithm}[H]
\caption{Stage I and stage II algorithms}
\begin{algorithmic}[1]
\STATE \textbf{-- STAGE I -- }
\STATE In the first cohort ($c_1 = 1$) patients 1 and 2 receive the dose combination $(x_1, y_1) = (x_2, y_2) = (0, 0)$
\STATE In the second cohort ($c_1 = 2$) patients 3 and 4 receive doses $(x_3, y_3)$ and $(x_4, y_4)$ respectively, where $y_3 = y_1$, $x_4 = x_2$, $x_3$ is the $\alpha$-th percentile of $\pi(\Gamma_{X|Y = y_1} | D_2)$, $y_4$ is the $\alpha$-th percentile of $\pi(\Gamma_{Y | X = x_2} | D_2)$.
\FOR{$c_1 = 3:C_1$}
  \IF{$c_1$ is an even number}
   \STATE Patient $2 c_{1} - 1$ receives the dose combination $(x_{2c_{1}-1}, y_{2c_{1}-3})$ where $x_{2c_{1}-1} = \Pi_{\Gamma_{X | Y = y_{2c_{1}-3}}}^{-1}(\alpha | D_{2c_{1}-2})$
   \STATE Patient $2c_{1}$ the dose combination $(x_{2c_{1}-2}, y_{2c_{1}})$, where $y_{2c_{1}} = \Pi_{\Gamma_{Y | X = x_{2c_{1}-2}}}^{-1}(\alpha | D_{2c_{1}-2})$
   \ELSE
   \STATE Patient $2c_{1} - 1$ receives the dose combination $(x_{2c_{1} - 3}, y_{2c_{1} - 1})$ where $y_{2c_{1} - 1} = \Pi_{\Gamma_{Y | X = x_{2c_{1}-3}}}^{-1}(\alpha | D_{2c_{1}-2})$
   \STATE Patient $2c_{1}$ receives the dose combination $(x_{2c_{1}}, y_{2c_{1} - 2})$ where $x_{2c_{1}} = \Pi_{\Gamma_{X | Y = y_{2c_{1}-2}}}^{-1}(\alpha | D_{2c_{1}-2})$
  \ENDIF
 \ENDFOR
 \STATE \textbf{-- STAGE II -- } 
 \STATE Calculate the posterior median of the parameters $(\widehat{\rho}_{00}, \widehat{\rho}_{10}, \widehat{\rho}_{01}, \widehat{\alpha}_3, \widehat{\beta}_0, \widehat{\beta}_{1}, \widehat{\beta}_{2}, \widehat{\beta}_{3}, \widehat{\beta}_{4}, \widehat{\beta}_{5})$ given data $D_{N_1}$
 \FOR{$c_2 = 1:C_2$}
    \STATE Calculate estimated MTD set $\widehat{\mathcal{M}} = \left \{ (x,y): y = \left ( \frac{F^{-1}(\theta_Z) - F^{-1}(\widehat{\rho}_{00}) - (F^{-1}(\widehat{\rho}_{10}) - F^{-1}(\widehat{\rho}_{00})) x}{(F^{-1}(\widehat{\rho}_{01}) - F^{-1}(\widehat{\rho}_{00})) + \widehat{\alpha}_3 x} \right ) \right \}$
    \STATE Generate a sample of dose combinations of size $m_2$ that belong to $\widehat{\mathcal{M}}$ from the (estimated) standardized density of $\widehat{\pi}_E(x,y)$, and assign it to the subsequent cohort of $m_2$ patients
    \STATE Calculate the posterior median of the parameters $(\widehat{\rho}_{00}, \widehat{\rho}_{10}, \widehat{\rho}_{01}, \widehat{\alpha}_3, \widehat{\beta}_0, \widehat{\beta}_{1}, \widehat{\beta}_{2}, \widehat{\beta}_{3}, \widehat{\beta}_{4}, \widehat{\beta}_{5})$ given data $D_{N_1 + c_2 \times m_2}$
\ENDFOR 
\end{algorithmic}
\label{algorithm1}
\end{algorithm}

\clearpage

Stage II can also be viewed as an extension of a response-adaptive randomization procedure with a finite number of doses (see \cite{berry2010bayesian}) to a response-adaptive randomization procedure with an infinite number of dose combinations. Mind that one could easily adapt Algorithm \ref{algorithm1} to have a single stage by starting directly with the adaptive randomization part of the design (i.e., stage II). The dose finding algorithm contains the following stopping rules for safety and futility:

\begin{itemize}
\item \textbf{Futility stopping rule}
    
For ethical considerations and to avoid exposing patients to sub-therapeutic dose combinations, we would stop the trial for futility if 
\begin{equation*}
    \underset{(x,y) \in \widehat{\mathcal{M}}_{n}}{\mbox{max}}\left [ P(F(\beta_{0} + \beta_1 x + \beta_2 y + \beta_3 xy + \beta_4 x^2 + \beta_5 y^2)  > \theta_E ~|~ D_n) \right ] < \delta_0,
\end{equation*}where $\delta_0$ is a pre-specified threshold. For the purpose of this article, we choose $\delta_0 = 0.1$. Mind that this stopping rule applies only after the initial cohort of patients in stage II.
    
\item \textbf{Safety stopping rule}

The design contains two stopping rules for safety, one for stage I and a different one for stage II. During stage I, we would stop the trial if
\begin{equation}
\label{eq_tox_stop_st1}
   P(\pi_Z(x = 0,y = 0) > (\theta + 0.1) ~|~ D_n) > \delta_{\theta_1},
\end{equation}where $\delta_{\theta_1} = 0.5$. In contrast, during stage II we would stop the trial if
\begin{equation}
\label{eq_tox_stop_st2}
   P(\Theta > (\theta + 0.1) ~|~ D_n) > \delta_{\theta_2},
\end{equation} where $\Theta$ represents the rate of DLTs for both stages of the design regardless of dose and $\delta_{\theta_2} = 0.7$ represents the confidence level (i.e., 70\%) that a prospective trial results in an excessive DLT rate. A non-informative Jeffrey's prior Beta$(0.5,0.5)$ is placed on the parameter $\Theta$. Mind that equation \eqref{eq_tox_stop_st2} does not depend on a pre-specified dose combination like equation \eqref{eq_tox_stop_st1}. The rational behind this choice is that it would be sub-optimal to condition the stage II safety stopping rule on a pre-specifed dose combination, even if selected at the end of stage I, given that, during stage II, the MTD curve is updated after each new cohort of patients.
\end{itemize}

\section{Simulation study}
\label{sc_simulation_study}

\subsection{Operating characteristics}
\label{sc_operating_characteristics}

Stages I and II are evaluated through the following operating characteristics:

\begin{enumerate}

    \item Pointwise average bias of the estimated MTD curve:
    
    Following \cite{tighiouart2017bayesian}, let $\mathcal{M}$ be the true MTD curve of a pre-specified scenario and $\widehat{\mathcal{M}}_j$ be the estimated MTD set at the $j$-th simulated trial, $j = 1, \dots, J$. The pointwise average bias when estimating the MTD is defined as follows. For every point $(x,y) \in \mathcal{M}$, let

    \begin{equation}
    \label{eq_pointwise_minimum_relative_distance}
    d_{(x,y)}^{(j)} = \mbox{sign}(y' - y) \min_{(x^*,y^*):(x^*,y^*) \in \widehat{\mathcal{M}}_j} \{ (x - x^*)^2 + (y - y^*)^2 \}^{1/2},
    \end{equation}where $y'$ is such that $(x,y') \in \widehat{\mathcal{M}}_j$. This is the minimum relative distance of the point $(x,y)$ of $\mathcal{M}$ to $\widehat{\mathcal{M}}_j$. The pointwise average bias is thus calculated as $d_{(x,y)} = \frac{1}{J} \sum_{j=1}^{J} d_{(x,y)}^{(j)}$.
    
    \item Percentage of correct MTD curve recommendation:
    
    Following \cite{tighiouart2017bayesian}, we define the percentage of correct MTD recommendation as the pointwise percentage of trial for which the minimum distance of the point $(x,y)$ on $\mathcal{M}$ to $\widehat{\mathcal{M}}_j$ is no more than $(100 \times p)\%$ of the true MTD. It is calculated using the definition of the minimum relative distance given by equation \eqref{eq_pointwise_minimum_relative_distance} and it is defined as

    \begin{equation*}
    R_{(x,y)} = \frac{1}{J} \sum_{j=1}^{J} I \{|d_{(x,y)}^{(j)} ~| \leq p \Delta (x,y) \},
    \end{equation*}where $0 < p < 1$, and $\Delta (x,y)$ represents the euclidean distance between the minimum dose combination $(0,0)$ and the point $(x,y)$ on the MTD curve. Intuitively, this metric is equivalent to drawing a circle with radius $p\Delta (x,y)$ and calculating the percentage of trials with $\widehat{\mathcal{M}}_j$ falling inside the circle, where $p$ is a tolerance parameter. 
    
    \item Average DLT rate and percentage of trials with DLT rate higher than $\theta_Z + 0.1$.
    
    \item Distribution of recommended optimal dose combinations.
    
    \item Proportion of recommended optimal dose combinations with true probability of efficacy above $\theta_E$.
    
    \item Bayesian power (type-I error probability under $H_0$):
    
    \begin{equation*}
    \label{eq_power_formula}
        \text{Power} \approx \frac{1}{J} \sum_{j = 1}^{J} \textbf{1} \left \{ \underset{(x,y) \in \widehat{\mathcal{M}}_N}{\mbox{max }} \left ( P(F(\beta_{0j} + \beta_{1j} x + \beta_{2j} y + \beta_{3j} xy + \beta_{4j} x^2 + \beta_{5j} y^2) > \theta_E | D_N  ) \right )   > \delta_u \right \},
    \end{equation*}where `$\textbf{1}(.)$' represents an indicator function.
    
    For scenarios favoring the null hypothesis as defined in \eqref{eq_hypotheses}, the above formula represents the Bayesian type-I error probability.
    
    \item Proportion of patients in stage II allocated to dose combinations with true probability of efficacy above $\theta_E$.
    
    \item Average posterior probability of early stopping for futility and safety.
    
\end{enumerate}

\subsection{Dose-toxicity and dose-efficacy scenarios}

\begin{figure}
\caption{True dose-toxicity profiles. The black dot in the dose combination of cisplatin and cabazitaxel 15/75 represents the MTD found by \cite{lockhart2014phase} using a '3+3' design.}
\centering
\includegraphics[scale=0.65]{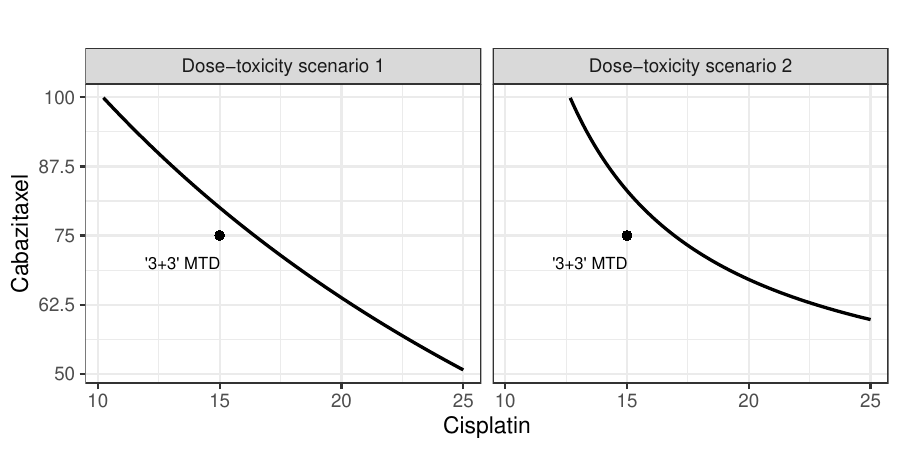}
\label{Figure_1}
\end{figure}

Following \cite{tighiouart2019two}, we use the two dose-toxicity scenarios that were expected by the principal investigator of the cisplatin-cabazitaxel trial, which are displayed in Figure \ref{Figure_1}. To generate dose-toxicity scenarios 1 and 2, we have used the parameter values $\{\rho_{00} = 1 \times e^{-7}, \rho_{01} = 0.3, \rho_{10} = 0.3, \alpha_3 = 2\}$ and $\{\rho_{00} = 1 \times e^{-5}, \rho_{01} = 0.01, \rho_{10} = 0.005, \alpha_3 = 9\}$, respectively. The target probability of DLT is $\theta = 0.33$.

\begin{figure}
\caption{True dose-efficacy profiles under the null and alternative hypotheses varying with the dose of Cisplatin along the true MTD curves. The red dotted line represents $\theta_E$ and the green dotted line represent the highest probability of efficacy.}
\centering
\includegraphics[scale=0.65]{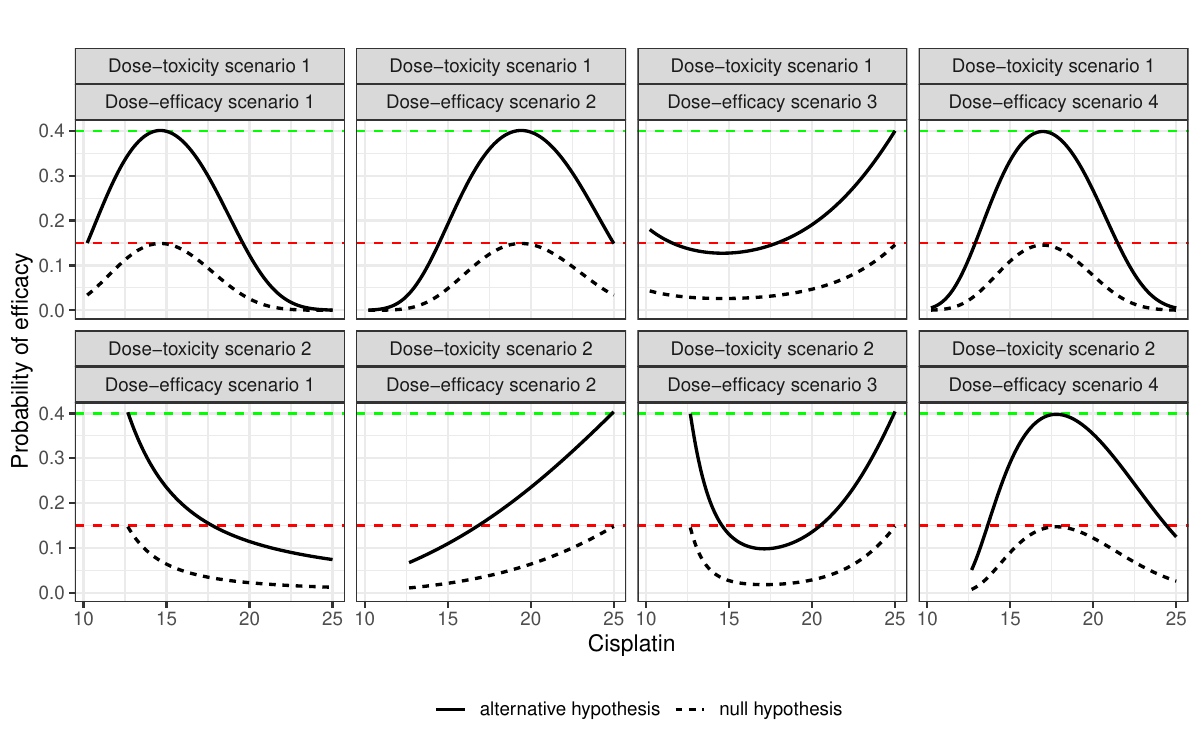}
\label{Figure_2}
\end{figure}

Within each of these dose-toxicity scenarios, we study four different dose-efficacy profiles, both under the alternative ($H_1$) and the null ($H_0$)  hypotheses, that place the most efficacious dose combination in a different region of the MTD curve. These profiles are obtained using model (\ref{eq_prob_eff}) for $(x,y)$ in the true MTD curve as defined by dose-toxicity scenarios 1 and 2. These are displayed in Figure \ref{Figure_2}. The parameter values used to generate the dose-efficacy profiles under $H_1$ and $H_0$ are displayed in Table \ref{table_dose_efficacy_values}. It is important to mention that all the dose-efficacy scenarios have been generated with the parameters $\beta_4, \beta_5 = 0$ (i.e., using monotonic dose-efficacy profiles) since we are combining two cytotoxic agents. These two parameters are incorporated for the sole purpose of adding some flexibility since, as mentioned in section \ref{sc_marginal_models}, in the development of this article we have found that, because of the assumed synergistic effect of the compounds in terms of the dose-toxicity profile, an efficacy model with only linear terms was not able to fully capture the dose-efficacy profile in one of the dose-efficacy scenarios we present (scenario 3).

The probability of response (clinical benefit) for the standard of care treatment is $\theta_E = 0.15$, and the cohort size in stage II is $m_2 = 5$. Under $H_1$, we assume an effect size of 0.25 (i.e., in all dose-efficacy profiles, the highest probability of efficacy is equal to $\theta_E + 0.25 = 0.4$). Under $H_0$, the highest probability of efficacy is equal to $\theta_E$. The sample sizes for Stages I and II are $N_1 = N_2 = 30$ and we simulated $J=1000$ trials using Algorithm 1. The DLT and efficacy responses were generated from models (\ref{eq_prob_dlt}) and (\ref{eq_prob_eff}), respectively.

\begin{table}
\caption{\label{table_dose_efficacy_values} Dose-efficacy parameters values used to generate the dose-efficacy profiles under $H_0$ and $H_1$ presented in Figure \ref{Figure_2}. The parameters $\beta_0 (H_0)$ and $\beta_0 (H_1)$ belong to the null and alternative hypotheses, respectively. The parameters $\beta_1, \beta_2, \beta_3, \beta_4, \beta_5$ are the same under the null and alternative hypotheses.}
\centering
\fbox{%
\begin{tabular}{|c|c|c|c|c|c|c|c|c|}
\hline
Dose-toxicity scenario & \multicolumn{4}{c|}{1} & \multicolumn{4}{c|}{2} \\ \hline
Dose-efficacy scenario & 1 & 2 & 3 & 4 & 1 & 2 & 3 & 4 \\ \hline
$\beta_0 (H_0)$ & -6.3 & -6.3 & -7.3 & -4.8 & -2.8 & -2.8 & -6.6 & -7.28 \\ \hline
$\beta_0 (H_1)$ & -5.51 & -5.51 & -6.5 & -4 & -2 & -2 & -5.8 & -6.49 \\ \hline
$\beta_1$ & 2 & 4.3 & 6.17 & 1.25 & 0.05 & 1.55 & 4.63 & 0.2 \\ \hline
$\beta_2$ & 4.3 & 2 & 5.5 & 1.25 & 1.57 & 0.05 & 4.73 & 0.2 \\ \hline
$\beta_3$ & 10 & 10 & 0 & 12 & 1 & 1 & 0 & 26 \\ \hline
$\beta_4$ & 0 & 0 & 0 & 0 & 0 & 0 & 0 & 0 \\ \hline
$\beta_5$ & 0 & 0 & 0 & 0 & 0 & 0 & 0 & 0 \\ \hline
\end{tabular}}
\end{table}

\subsubsection{Results}

\begin{figure}[t]
\caption{Pointwise average bias of the estimated MTD curve with respect to the true MTD curve under the null (dashed line) and alternative (solid line) hypotheses, respectively. Results are displayed from the point of view of the standardized Cisplatin.}
\centering
\vspace{0.25cm}
\includegraphics[scale=0.65]{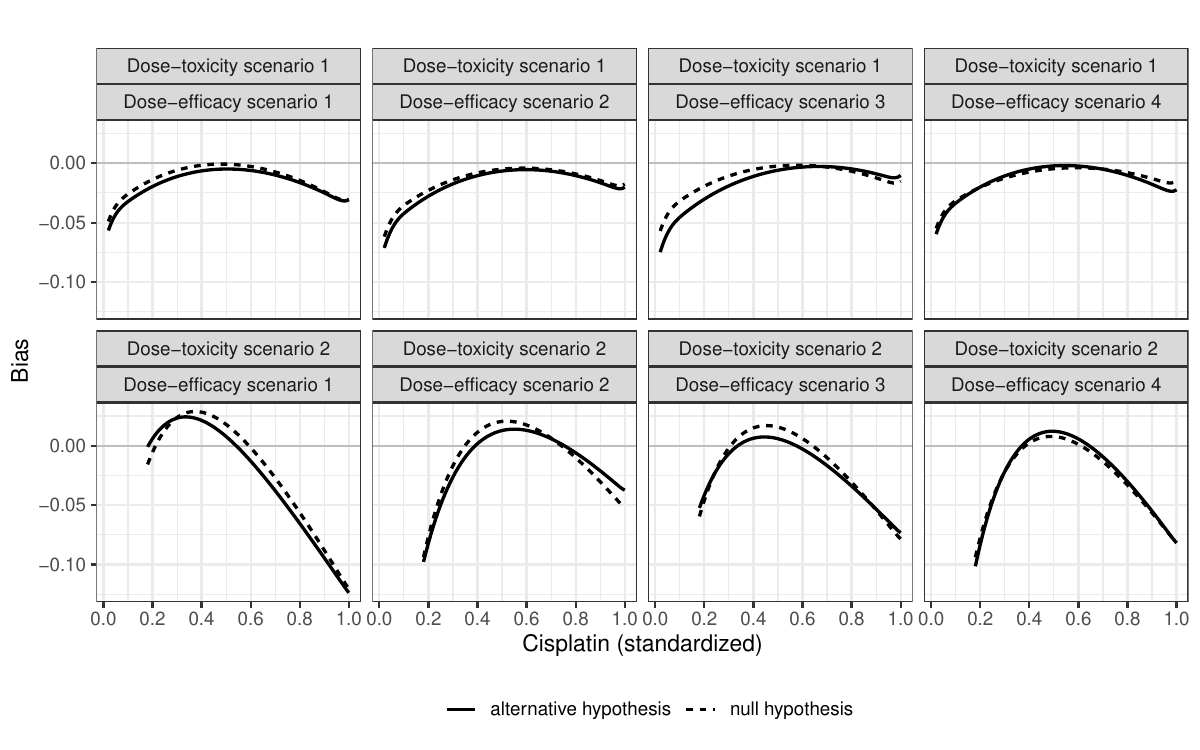}
\label{plot_average_pointwise_MTD_bias}
\end{figure}

In Figure \ref{plot_average_pointwise_MTD_bias}, we present the pointwise average bias of the estimated MTD curve at the end of the trial with respect to the true MTD curve under the null (dashed line) and alternative (solid line) hypotheses, respectively. In line with the results presented by \cite{tighiouart2017bayesian,jimenez2020bayesian}, the design yields a good estimation of the MTD curve, specially around its central region whereas this estimation is less precise at the extremes of the MTD curve. However, because of the use of all toxicity data from stages I and II, the average bias we observe in the MTD curve estimation is, overall, lower than the values reported in the supplementary material of \cite{jimenez2020bayesian} with also vague prior distributions. In Figures S1 and S2-S3 in the supplementary material, we display the estimated MTD curves and the percentage of correct MTD recommendation under the null and alternative hypotheses, respectively. In most scenarios, the pointwise percent selection varies between $60 \%$ and $90\%$ except near the edges of the true MTD curve using a threshold value of $p=0.1$ and between $80 \%$ and $100\%$ using a tolerance threshold $p=0.2$. These results are consistent with the conclusions we draw from the pointwise average bias.

In Table \ref{table_dlt_rate} we present the safety operating characteristics and we observe that, on average, the DLT rate is between 29 - 33\% and the percentage of trials with proportion of DLT above $\theta_Z + 0.1$ ranges between 0 - 0.2\%. Therefore, the design is considered safe. 

\begin{table}
\caption{\label{table_dlt_rate}Average DLT rate and percentage of trials with DLT above $\theta_Z + 0.1$ .}
\centering
\fbox{%
\begin{tabular}{|c|c|c|c|c|c|c|c|c|}
\hline
Dose-toxicity scenario & \multicolumn{4}{c|}{1} & \multicolumn{4}{c|}{2} \\ \hline
Dose-efficacy scenario & 1 & 2 & 3 & 4 & 1 & 2 & 3 & 4 \\ \hline
Average DLT rate $(H_0)$ & 0.333 & 0.337 & 0.333 & 0.336 & 0.289 & 0.304 & 0.299 & 0.330 \\ \hline
Average DLT rate $(H_1)$ & 0.336 & 0.339 & 0.337 & 0.337 & 0.289 & 0.308 & 0.300 & 0.332 \\ \hline
\makecell{Percentage (\%) of trial \\ with DLT rate above \\ $\theta_Z + 0.1$ $(H_0)$} & 0.0 & 0.1 & 0.2 & 0.1 & 0.0 & 0.0 & 0.0 & 0.1 \\ \hline
\makecell{Percentage (\%) of trial \\ with DLT rate above \\ $\theta_Z + 0.1$ $(H_1)$} & 0.1 & 0.1 & 0.0 & 0.0 & 0.0 & 0.0 & 0.0 & 0.1 \\ \hline
\end{tabular}}
\end{table}

\begin{figure}[t]
\caption{Bayesian power and type-I error probability for scenarios under the alternative and null hypotheses, respectively.}
\centering
\vspace{0.25cm}
\includegraphics[scale=0.7]{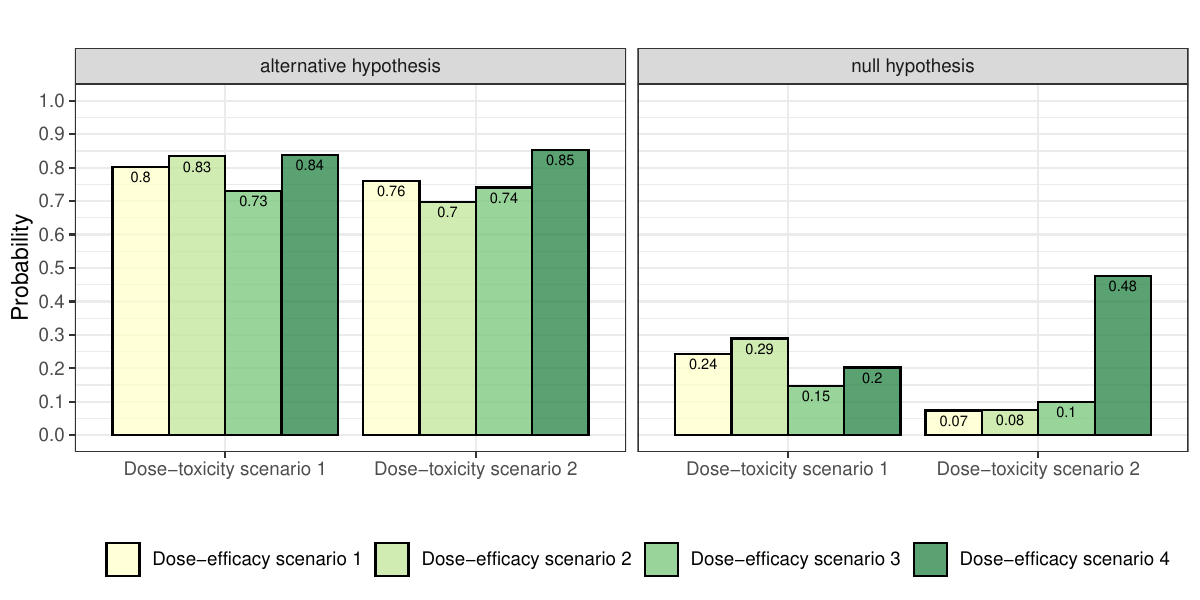}
\label{plot_power}
\end{figure}

In Figure \ref{plot_power}, we show the probability of rejecting the null hypothesis, which corresponds to the (Bayesian) power under the alternative hypothesis and to the (Bayesian) type-I error probability under the null hypothesis. This probability ranges between 0.7 and 0.85 under the alternative hypothesis (i.e., power), and between 0.07 and 0.29 under the null hypothesis for all but one scenario (i.e., type-I error). These values are in line with the results presented by Tighiouart (2019) \cite{tighiouart2019two} and Jim\'enez et al. (2020) \cite{jimenez2020bayesian}. We observe one interesting result under the null hypothesis in dose-toxicity scenario 2 - dose-efficacy scenario 4, where the type-I error probability is close to 0.5. This value is a consequence of a slight overestimation of the MTD curve in its central region (see Figure \ref{plot_average_pointwise_MTD_bias}). This behavior is not present in dose-toxicity scenario 1 - dose-efficacy scenario 4 where the MTD curve is practically unbiased in its central region. By overestimating the MTD, we are no longer having a maximum probability of 0.4 and 0.15 under the alternative and null hypotheses, respectively. This causes an increase of both the power and the type-I error probability. In fact, dose-toxicity scenario 2, dose-efficacy scenario 4 is the scenario that achieves both the highest power and highest type-I error probability because of the overestimation of the MTD curve in the region of highest efficacy. By the same token, an underestimation of the MTD curve in the region of highest efficacy will lead to both lower power and lower type-I error probability. 

\begin{figure}[t]
\caption{Density of the dose combinations recommended at the end of each simulated trial as the optimal dose combination under the alternative hypothesis. The solid black line represents the true MTD curve of each dose-toxicity scenario and the red dot represents the true optimal dose combination.}
\centering
\vspace{0.25cm}
\includegraphics[scale=0.7]{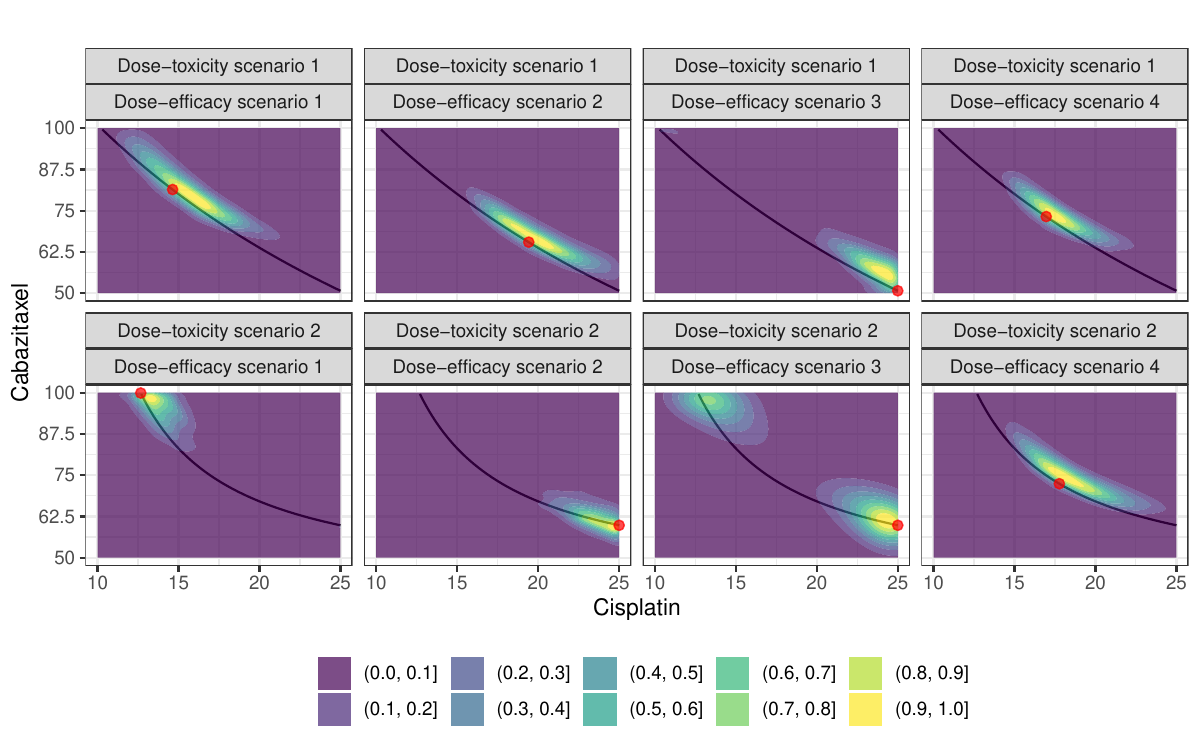}
\label{plot_optimal_dose_combination_H1}
\end{figure}

In Figure \ref{plot_optimal_dose_combination_H1}, we present the density (i.e., the distribution) of the dose combinations recommended at the end of each simulated trial as the optimal dose combination. In this article, the MTD curve is never fixed which translates into recommended optimal dose combinations above and below the true MTD curve. However, we can see how the density of these recommended optimal dose combinations is, generally speaking, very close to the true MTD curve, with the region of higher mass (i.e., bright yellow) containing the true optimal dose combination. Mind that in dose-toxicity scenario 2 - dose-efficacy scenario 3, the true probability of efficacy at the left extreme of the true MTD curve is intended to be very close to the one from the optimal dose combination located at the right extreme of the MTD curve (see Figure \ref{Figure_2}). The bi-modality observed in the density of the recommended optimal dose combinations in this scenario shows that the design is able to capture the characteristic of this dose-efficacy scenario.

In Figure S4 (supplementary material), we present the distribution of dose combinations allocated to all patients during stage II, which works under the adaptive randomization principle. We display all scenarios under the alternative hypothesis only. Results clearly show how the mode of the distribution corresponds to the dose combination with highest probability of efficacy under each scenario. Moreover, the proportion of patients during stage II that were allocated to dose combinations with a true probability of efficacy above $\theta_E$ ranges between 69-77\% (see Figure S5 in the supplementary material).This result also holds under the null hypothesis (results not shown).  

In Figure S6, we present the average posterior probability of early stopping for the stopping rules we pre-specified in section \ref{sc_study_design}. Under the alternative hypothesis, we see that the probability of early stopping for futility is always below 0.02. Under the null hypothesis, this probability ranges generally between 0.067 and 0.27, depending on the scenario. We see how the probability of early stopping for futility in dose-toxicity scenario 2 - dose-efficacy scenario 4 is much lower (0.067) than in other scenarios because of the overestimation of the central region of the MTD curve and the allocation of patients with higher probability of efficacy. In terms of the probability of early stopping for safety, we see that under both the null and alternative hypotheses, this probability is practically equal to 0. The values presented here are in line with the operating characteristics from \cite{tighiouart2019two}.

\subsubsection{Model Robustness}

In Figure S9 of the supplementary material, we present the power and type-I error obtained under model misspecification. In this assessment, the true underlying dose-toxicity and dose-efficacy relationships' link functions are different from the probit link function used in the design's working marginal models (see section \ref{sc_marginal_models}). In simple terms, toxicity and efficacy data will be generated using a common link function (e.g., logistic) and the design will proceed under the (wrong) assumption that such common link function is a probit link function. The purpose of this robustness assessment is to test whether the design is able to maintain the operating characteristics presented in previous sections under small deviations between the true underlying marginal models and the working marginal models. To limit the number of comparisons, the toxicity and efficacy marginal models employ the same link function.

For the assessment, we reproduced the first dose-toxicity scenario presented in Figure \ref{Figure_1} and the first two dose-efficacy scenarios presented in Figure \ref{Figure_2} using the logistic and the complementary log-log link functions. Because the link functions are different, the dose-toxicity and dose-efficacy surfaces have to be different, specially around the extremes of the the surfaces. We refer the reader to \cite{tighiouart2014dose} to see how the dose-toxicity surfaces differ using different link functions. Under model misspecification, the true models' parameters are selected in a way that there is an overlap of the toxicity and efficacy profiles of all link functions in the region where the optimal dose combination is located. In Figures S7 and S8 of the supplementary material we display the MTD curves as well as the efficacy profiles on the MTD curves generated with the logistic and complementary log-log link functions. 

The results presented in Figure S9 show that our design is able to maintain its performance under mode misspecification, with power and type-I error values with minor differences between the three tested link functions. This is not unexpected since the quadratic terms included in the dose-efficacy model allow for higher flexibility.

Last, and for illustration purposes, in Figures S10 and S11 of the supplementary material we present the contours of the marginal toxicity and efficacy models, respectively, with the logit, logistic and complementary log-log link functions. For this illustration we selected dose-toxicity scenario 1 and dose-efficacy scenario under the alternative hypothesis. In this robustness assessment, the true toxicity and efficacy model parameters are selected so that the MTD curves and the efficacy profiles within the MTD curves match across all the link functions (see Figures S7 and S8). However, outside of the MTD curves, we observe how each link function induces slightly different dose-toxicity and dose-efficacy surfaces.

\section{Discussion}
\label{sc_discussion}

We present a Bayesian two-stage design for cancer phase I-II clinical trials combining two cytotoxic agents with continuous dose levels. Motivated by the cisplatin-cabazitaxel trial, we extend the work of \cite{tighiouart2019two} by allowing the continuous update of the maximum tolerated dose (MTD) curve throughout the entire study as well as the use of stage I efficacy data under the assumption of a single patient population across stages.
In stage I we aim to estimate the MTD curve following the escalation with overdose control (EWOC) principle and enrolling up to 30 patients. In stage II we model treatment efficacy as a binary indicator of treatment response using a probit quadratic model, and we enrol 30 additional patients using response adaptive randomization. The rational for quadratic terms in a setting of two cytotoxic agents is that, under the presence of (synergistic) interaction between the compounds, the dose-efficacy profile along the MTD can become quite complex, and we found that a linear model was not able to correctly identify the region with the most efficacious dose combinations in certain scenarios. However, under the presence of non-interactive compounds the dose-efficacy profiles cannot surpass a certain level of complexity and thus a linear dose-efficacy model would be sufficient to identify the most efficacious dose combinations.

We study the properties of our design under the two (synergistic) dose-toxicity scenarios that were expected by the principal investigator of the cisplatin-cabazitaxel trial introduced by \cite{tighiouart2019two} with 30 patients in each stage and an effect size of 0.25 in dose-toxicity scenarios under the alternative hypothesis. We study four different dose-efficacy profiles that place the most efficacious dose combination in different regions of the MTD curve (i.e., left extreme, right extreme, central and both extremes).
The estimated MTD curve tends to be slightly biased around its extremes and less biased around its central region as reported by \cite{tighiouart2017bayesian,jimenez2020bayesian}. In fact, because of the continuous update of the MTD curve during stage II, we have observed a slight reduction of the overall estimation bias with respect to the results presented by \cite{jimenez2020bayesian} where the MTD curve was only updated in stage I.
The proposed probit quadratic efficacy model has a good performance and is able to fully capture the different dose-efficacy profiles along the true MTD curve. This is fully reflected in Figure \ref{plot_optimal_dose_combination_H1}, dose-toxicity scenario 2 - dose-efficacy scenario 3, where the probability of efficacy if high at both extremes of the true MTD curve. As a matter of fact, in the development of this research, we found that a marginal efficacy model with only linear terms was not able to capture this particular profile which is heavily conditioned by the synergistic nature of the MTD curve. Thus, adding quadratic terms allows to better capture any dose-efficacy profile along any MTD curve.
Overall, the design is safe and yields good operating characteristics across scenarios with power and type-I error rates that are consistent with the values reported by \cite{tighiouart2019two,jimenez2020bayesian}.
Another contribution of this article is that we account for the uncertainty of the estimated MTD curve during stage II since we continuously update it after each cohort of patients. In fact, we have observed its behavior in stage II, which was not explored in \cite{tighiouart2019two,jimenez2020bayesian}. Results show that overall bias in the MTD curve is also driven by the adaptive randomization part of the design where we allocate patients to dose combination that are likely to be efficacious along the estimated MTD curve. This type of patient allocation unbalances the amount of data that we have in different regions of the estimated MTD curve, which could potentially damage the precision of the estimation in the search of higher efficacy. We believe these findings are coherent with the fact that we jointly work with i) the uncertainty around the MTD curve estimation and ii) the uncertainty around the estimation of the dose-efficacy profile. With these conclusions in mind, we strongly recommend using informative prior distributions whenever possible, and/or a slightly more conservative EWOC criterion (i.e., a lower percentile) either throughout the entire study or in stage II only. An alternative approach to hypothesis testing is to report the percent selection of correct optimal dose combinations, as is traditionally done in phase I/II dose finding studies. Since we are dealing with continuous dose levels, a reasonable definition of this metric would be the percentage of recommended optimal dose combination with (true) $\pi_Z(x,y) \leq \theta_Z + \delta_1$ and (true) $\pi_E(x,y) \geq \theta_E^* - \delta_2$, where $\theta_E^*$ represents the true maximum probability of efficacy of a given scenario (i.e., $\theta_E^* = 0.4$ and $\theta_E^* = 0.15$ in the dose-efficacy scenarios that we considered) and $\delta_1, \delta_2$ are pre-specified positive threshold values.

In this article, we have not used the informative prior distributions used by \cite{tighiouart2019two} and therefore our estimation of the MTD curve is consequently less precise. The operating characteristics presented in this article can be considered as the baseline operating characteristics this design can have given the use of vague prior distributions and relatively low sample size. By increasing the sample size or using more informative prior distributions in the dose-toxicity model, the design's operating characteristics are expected to improve. 

Finally, model robustness was explored by using logistic and complementary log-log link functions to generate toxicity and efficacy data. Results show that the proposed design is robust to model misspecification under small deviations of the true model from the working model.

On the application of the proposed design to settings with other type of compounds (i.e., immunotherapies or targeted therapies), we note that major changes would be needed to adapt the design to the characteristics of such compounds. The most relevant adaption would be the definition of the optimal dose combination combination. With cytotoxic agents, it is generally assumed that the optimal dose combination will be somewhere along the MTD curve. With immunotherapies or targeted therapies, this is not guaranteed and utility functions become necessary to find the right balance between toxicity and efficacy (see e.g., \cite{liu2018bayesian}). Also, because we would no longer restrict the search to the estimated MTD, we would need to extend the stage II adaptive randomization algorithm so that we are allowed to allocate patients to any dose combination with estimated probability of toxicity below the target probability of toxicity. 

Also, the design could be implemented in a setting with discrete dose combinations either with cytotoxics or immunotherapies with minor modifications with respect to the version of the design with continuous dose levels. To do so, in stage I, the continuous dose combinations would be rounded to nearest discrete dose combinations whereas, in stage II, patients would only be allocated to the initial set of discrete combinations based on the updated probability of efficacy estimates (i.e., the higher estimated probability of efficacy of a dose combination the higher the probability of being allocated to that dose combination).

These two extensions are part of our research plan and we intent to evaluate the operating characteristics of the proposed design under these settings.

\subsection*{Data availability}

The \texttt{R} and \texttt{JAGS} scripts necessary to fully reproduce the results presented in this article are available at \newline
\href{https://github.com/jjimenezm1989/Phase-I-II-design-combining-two-cytotoxics}{\small{\texttt{https://github.com/jjimenezm1989/Phase-I-II-design-combining-two-cytotoxics}}}

\subsection*{Funding}

Mourad Tighiouart is funded by NIH the National Center for Advancing Translational Sciences (NCATS) UCLA CTSI (UL1 TR001881-01) and NCI P01 CA233452-02.

\subsection*{Disclaimer}

Jos\'e L. Jim\'enez is employed by Novartis Pharma A.G. who provided support in the form of salary for the author, but did not have any additional role in the preparation of the manuscript. Also, the views expressed in this publication are those of the authors and should not be attributed to any of the funding institutions or organisations to which the authors are affiliated.

\bibliographystyle{plain}
\bibliography{references.bib}

\end{document}